\documentstyle[preprint,eqsecnum,aps,epsf]{revtex}

\begin{document}
\draft
\preprint{}
\title{Enhancement of the electronic contribution \\to the low temperature specific heat\\
of Fe/Cr magnetic multilayer
}
\author{B. Revaz, M.-C. Cyrille, B. Zink, Ivan K. Schuller and F. Hellman}
\address{
Dept of Physics, University of California at San Diego\\
9500 Gilman Dr, La Jolla, CA 92093
}

\date{\today}
\maketitle
\begin{abstract}

We measured the low temperature specific heat of a sputtered $(Fe_{23\AA}/Cr_{12\AA})_{33}$ magnetic multilayer, as well as separate $1000 \;\AA$ thick Fe and Cr films. 
Magnetoresistance and magnetization measurements on the multilayer demonstrated antiparallel coupling between the Fe layers.
Using microcalorimeters made in our group, we measured the specific heat for $4<T<30 \; K$ and in magnetic fields up to $8 \; T$ for the multilayer.
The low temperature electronic specific heat coefficient of the multilayer in the temperature range $4<T<14 \; K$ is $\gamma_{ML}=8.4 \; mJ/K^{2}g-at$.
This is significantly larger than that measured for the Fe or Cr films ($5.4$ and. $3.5 \;mJ/K^{2}mol$ respectively). No magnetic field dependence of $\gamma_{ML}$
 was observed up to $8 \; T$. These results can be explained by a softening of the phonon modes observed in the same data and the presence of an Fe-Cr alloy phase at the interfaces.

\end{abstract}

\narrowtext
\section{Introduction}

Magnetic mutlilayers (MML), heterostructures of alternating ferromagnetic layers and non-magnetic spacers, have attracted attention in the past 10 years because of their implications both for fundamental research and technical applications. The most remarkable properties of these materials are the antiparallel coupling of the ferromagnetic layers for particular thickness of the non-magnetic spacer and the giant magnetoresistance (GMR) measured between this antiparallel configuration and the parallel one obtained by application of a magnetic field \cite{baibich}, \cite{schuller99}.
This latter property is of direct interest for magnetic sensor technology and has been successfully implemented in a wide range of applications \cite{dieny}. However, while the GMR is widely accepted as being due to spin dependent scattering, the details of the mechanism are still the subject of intense investigations. In particular, a detailed description of the GMR in terms of the local electronic structure is still not available.

Fe/Cr MML present an additional interest because of the electronic properties of the Cr spacer. Bulk Cr is known to be an incommensurate spin density wave antiferromagnet ($T_N=311\;K$), which is a consequence of the nesting of its Fermi surface (for a review on bulk Cr see \cite {fawcett88}). The Fermi surface of Cr has been the subject of particular attention because of its similarity to that of high temperature superconductors \cite {ruvalds}. In thin films, the magnetic properties of Cr are strongly modified by the stress and the presence of ferromagnetic capping layers as observed by neutron diffraction \cite {bodeker99}. In Fe/Cr (001) epitaxial MML, Fullerton et al. \cite {fullerton96} showed the loss of magnetic long range order in Cr layers for thicknesses $t_{Cr}<42\;\AA$. The electronic structure of these thin non-magnetic Cr layers is not well documented in spite of its importance for the calculations of the GMR. Beside the disappearance of the long range magnetic order that is attributed to the frustration of the Cr magnetic moments at the interfaces \cite {bodeker98}, it is expected that the Fermi surface of Cr is affected by exchange interactions with magnetic Fe. Angle resolved photoemission spectroscopy (ARPES) made on a single Cr layer epitaxially grown on atomically flat Fe shows interesting features such as the presence of spin polarized electronic quantum well states \cite {himpsel99}. However, this measurement cannot be directly compared with MML because the boundary conditions of the Cr upper layer are not the same as those of a Cr layer located deeper in the MML that is sandwiched between two Fe layers (for a discussion on the boundaries condition in the ARPES measurement see Himpsel et al. \cite {himpsel98} p 567). In addition, GMR is also observed in polycrystalline samples with rougher interfaces, for which the ARPES result may not be relevant.

Low temperature specific heat is a useful complementary technique to ARPES, since it probes the electronic density of states (DOS) at the Fermi level $N(E_F)$, a quantity that controls transport phenomena and itinerant magnetism.  Moreover the specific heat is sensitive to both the surfaces and the bulk of the sample. Recent progress in microcalorimetry \cite{denlinger} opened access to investigation of thin films and microstructures with thickness of $1000\;\AA$, which was prevented in the past by the small masses of these samples. We present in this paper low temperature specific heat measurements of an Fe/Cr MML. The Debye temperature and electronic coefficient are extracted for the first time for $1000\;\AA$ films. This paper is divided in two parts: First, we show in some detail how electronic terms are extracted from the specific heat of metallic films of $1000\;\AA$ thickness. Second, we present specific heat data of a sputtered Fe/Cr MML and compare it to that of Fe and Cr films. Our results show that the electronic contribution of the Fe/Cr MML is different from that expected in a simple model where bulk values of Fe and Cr are used. Possible origins for this result are discussed.

\section{Experimental details}
(110) textured Fe/Cr superlattices were directly sputtered  at room temperature on the membrane of a microcalorimeter. The microcalorimeters consist of a $1\;cm \times 1\;cm$ Si (100) frame surrounding a $0.5\;cm \times 0.5\;cm$ window covered with a low stress $a\!-\!Si_{1-x}N_{x}$ $1800 \; \AA$ thick membrane \cite {denlinger}. The film is sputtered through a $0.25\;cm \times 0.25\;cm$ shadow mask so that the deposited sample, located at the center of the membrane, is thermally insulated from the Si frame. For characterization purposes, a Si (100) substrate covered with $a\!-\!Si_{1-x}N_{x}$ has been placed next to the microcalorimeter during the deposition to serve as a test sample. 33 Fe/Cr bilayers have been deposited at an averaged rate of $2\;\AA/s$ with a base pressure of $6\!\times \!10^{-7} \;torr$ that was increased to $5\!\times \!10^{-3} \;torr$ of Ar during the deposition. We briefly give a description of the characterization of the MML. A detailed description of Fe/Cr MML deposited in our laboratory is reported elsewhere \cite {cyrille} and we refer the reader to this study for more details on the microstructure.

The thickness of the Cr layer was adjusted to obtain antiparallel coupling of the Fe layers. The thickness of the film and bilayers, obtained by fitting XRD data made on the test sample, are $1159\;\AA$ and $35\;\AA$ respectively. The thickness of the Fe layers is $23\;\AA$ as measured from the total thickness of an Fe film deposited under the same conditions as the MML without Cr, implying a Cr layer thickness of $12\;\AA$. Magnetization and magnetotransport measurements were performed on the test sample. Magnetization made at $10\;K$ with the magnetic field parallel to the layers showed a saturation magnetization 0 to the volume of the Fe layers of $M_s=1195\;emu/cm^3$ to be compared to $1710 \;emu/cm^3$ for bulk Fe. A small hysteresis is present, 0 by a remanent field of $H_r=300\;Oe$ and a remanent magnetization $M_r/M_s=0.23$. Magnetotransport measured at $10\;K$ with the van der Pauw method with current in the plane gave a residual resistivity $\rho_\circ=37\;\mu\Omega cm$ and a magnetoresistance $\Delta R/R(0)=16.8\%$, with a saturation field of $10\;kOe$. These data confirm a predominant antiparallel coupling between the Fe layers expected for $t_{Cr}=12\;\AA$, in agreement with other studies on Fe/Cr MML (see for example \cite {gijs92}, \cite{cyrille}). The non-vanishing remanent magnetization is attributed to pinholes that induce ferromagnetic shorts or local change of coupling due to Cr thickness fluctuations. Interdiffusion at the interfaces reduces the saturation magnetization of the sample with respect to the value expected for bulk Fe. These results are in agreement with those reported for MML with the same number of bilayers and deposited with the same Ar pressure \cite{cyrille}.

Two additional microcalorimeters, one with an Fe and the other with a Cr film, were prepared under the same deposition conditions. All microcalorimeters used in this work were chosen from the same wafer, which keeps variations in the microcalorimeter properties to less than 5 \% \cite{denlinger}. The thicknesses of the Fe  ($1050 \; \AA$) and Cr films ($1035 \; \AA$) are close to that of the MML to allow direct comparison of the raw data.

The relaxation method \cite {bachmann} was used to measure the specific heat. Application of this technique to the $a\!-\!Si_{x}N_{1-x}$ microcalorimeters is described in detail elsewhere \cite {denlinger}. This technique is especially suitable when absolute values of the specific heat are desirable over a large temperature range. When the internal thermal time constant $\tau_{int}$ is much faster than the external time constant $\tau_{e}$, the dynamics of the 2D heat transfer reduces to a 1D single time constant relaxation. The following simple equation, then, relates the measured $\tau_e(T)$ and thermal conductance linking sample and Si frame $\kappa(T)$ to the total specific heat $c_{tot}(T)$:

\begin{equation}
c_{tot}(T)={\tau_e(T) \times \kappa (T)}
\label{one}.
\end{equation}

Contributions to $c_{tot}$ include the heat capacities of the sample $c_{sam}$ and the addenda $c_{add}$ that consists of the thermometers and heater, the membrane beneath the sample and part of the thermal link. If a thermal conduction layer is necessary, as will be discussed in more detail below, its heat capacity adds to $c_{add}$. For our microcalorimeters, the thermal link at low temperature consists of the membrane located between the sample and the frame and 8 Pt leads ($0.25\times0.005\times 5\!\times\!10^{-6}\;cm^3$). Using numerical simulation of the 2D heat flow of our microcalorimeter, we obtain the result that 22.3\% of the thermal link has to be included in $c_{add}$ \cite{revaz}. This is less than the 33\% calculated by  Bachmann et al. \cite{bachmann} in the 1D case. The same simulations showed that this contribution does not depend on $c_{sam}$ when parameters (density, thickness and specific heat) of the samples measured in this study are used. In that case, this contribution is cancelled out in the difference $c_{sam}=c_{tot}-c_{add}$, which justify this subtraction technique to extract the specific heat of the sample.

A necessary and testable condition for the validity of Eq. \ref{one} is a large difference between the thermal conductances of the sample and the membrane ($\tau_{int}<<\tau_{e}$). In that case, the temperature is constant across the sample and drops between the edge of the sample area and the frame so that the measured $\kappa(T)$ is dependent solely on the thermal link. Under standard procedures, a thermal conduction layer is deposited either over or under the sample to ensure this. To avoid an unnecessary increase of $c_{add}$ that results from the thermal conduction layer, we first checked if the thermal conductivity of the Fe, Cr or Fe/Cr samples is sufficient to ensure a good thermal homogeneity across the sample area.
Estimates of the temperature gradient across the sample can be made since two $Nb\!-\!Si$ thermometers with different geometries are located on the back of the film. Steady state measurements performed on the Cr film at $18.72\;K$ with a power of $1\;\mu W$ applied to the Pt sample heater showed an increase of $\Delta T=0.54\;K$ in the first thermometer and $0.71\;K$ in the second (see Figure 1 of Ref. \cite{denlinger} for details concerning the design of the microcalorimeters). This temperature difference is due to the low thermal conductivity of the Cr film. As a consequence, single time constant measurements (Eq. \ref{one}) cannot be made on the microcalorimeter loaded with Cr only. The same conclusion holds for the Fe and Fe/Cr MML as similar or lower thermal conductivities are expected for these samples.

To improve the thermal conductivity of the samples, a Cu layer was deposited on top of the samples: The three microcalorimeters together with a bare microcalorimeter taken from the same wafer were put next to each other in a resistive evaporation chamber in which Cu  was evaporated. The thickness of the Cu layer ($1810\;\AA$) was measured using a Dektak profilometer on test samples that had been placed close to the microcalorimeters during the deposition. The dispersion of the thickness was found to be about $50\;\AA$ which is the (absolute) precision of the profilometer. The same measurement of temperature homogeneity was performed on the Cr sample but now with the Cu thermal conduction layer in the temperature range $10\!<\!T\!<\!30\;K$ where both thermometers can be measured : No difference of temperature was measured between the two thermometers within the 0.3\% error bars of the temperature measurement.

\begin{figure}
\centerline{\epsfbox{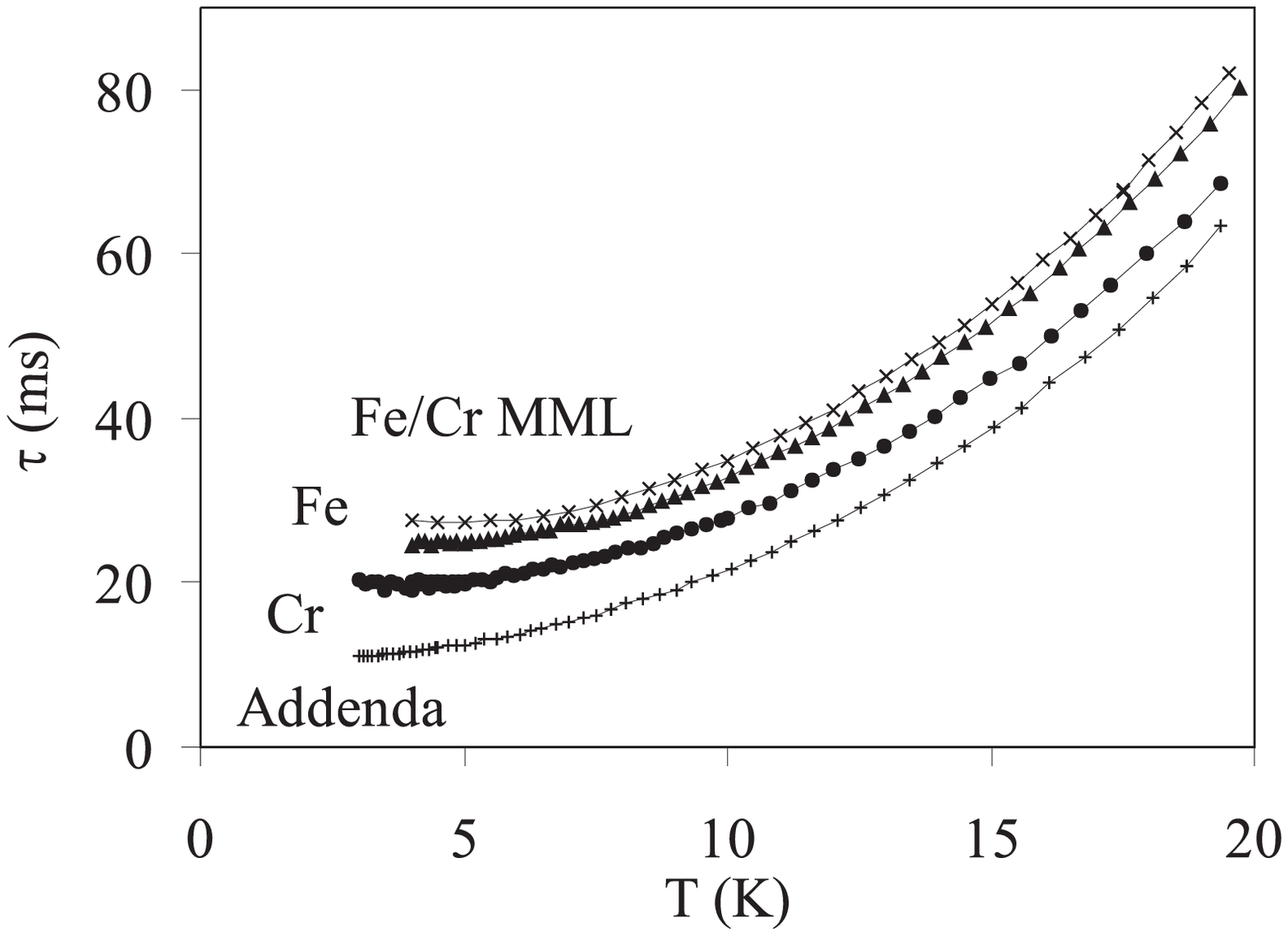}}
\caption{Experimental relaxation time constants of the bare addenda (with Cu thermal conduction layer only) ($+$), Cr ($\bullet$), Fe films ($\triangle$) and Fe/Cr MML ($\times$). These values are obtained by fitting the temperature decays of the sample thermometer when the current flowing in the heater is switched off.}
\label{figtau}
\end{figure}

The time constant $\tau_e$ is obtained by recording the off-balance signal of an ac resistive bridge - of which one arm is the NbSi thermometer - when the current applied to the Pt heater is switched off. The amplitude of the temperature change is set to 1 \% of the baseline temperature $T_\circ$. 20 decays are averaged and fitted using a single exponential function with a typical error of 0.8\%. Figure \ref{figtau} shows the measured $\tau_e$ for the addenda, Cr, Fe and Fe/Cr samples in the temperature range $3\!<\!T\!<\!20\;K$.

The integrated thermal conductance $\kappa (T_\circ,\Delta T)$ is obtained by applying a known power $P$ in the heater and measuring the steady state increase of temperature of the sample $\Delta T$ which gives:
\begin{equation}
\kappa (T_\circ,\Delta T)={{\int_{T_\circ}^{T\circ+\Delta T} \kappa(T')dT'}\over{\Delta T}}={P \over \Delta T}.
\label{two}
\end{equation}

\begin{figure}
\centerline{\epsfbox{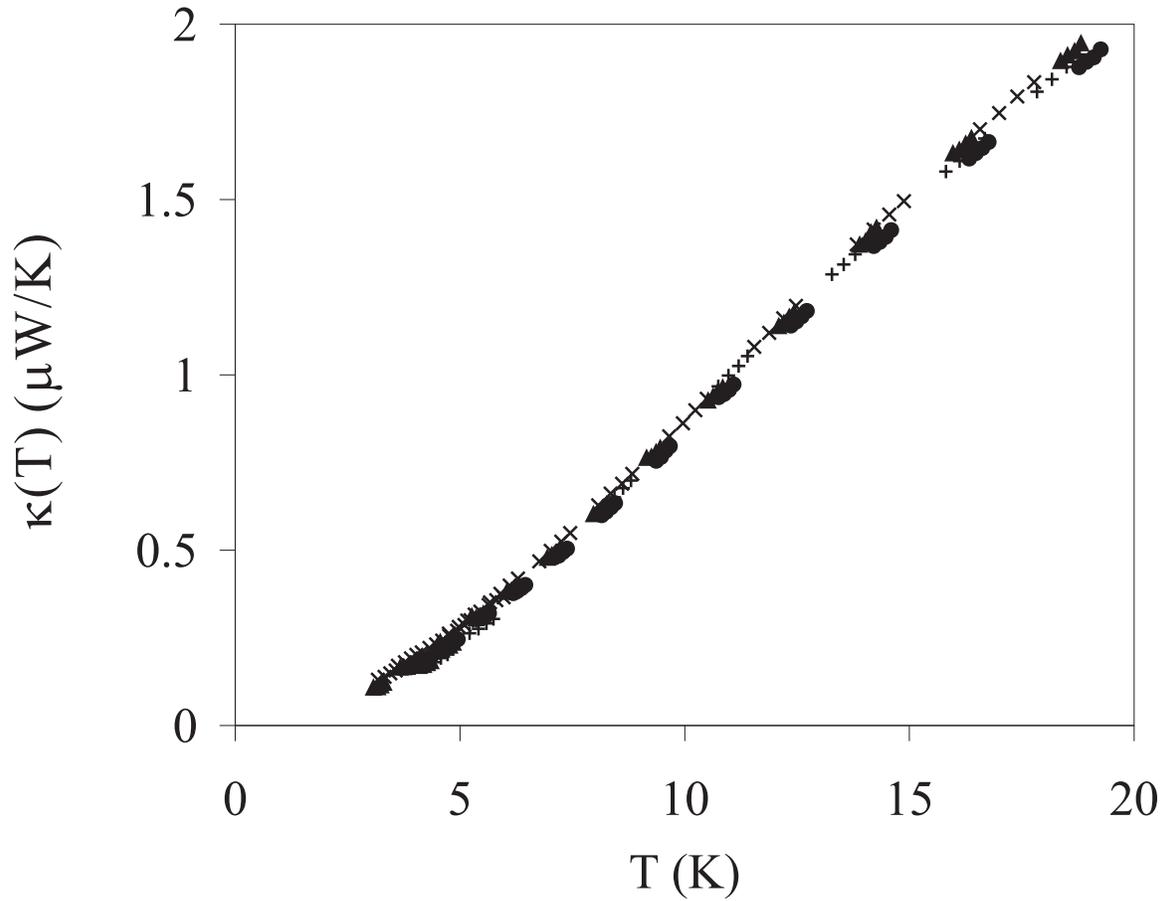}}
\caption{Integrated conductance of the membrane $\kappa(T_\circ+\Delta T/2)\approx \kappa(T_\circ,\Delta T)$ as a function of the mean temperature $T\!=\!T_\circ+\Delta T/2$ of the samples Fe/Cr ($\times$), Fe ($\triangle$), Cr ($\bullet$)  and addenda ($+$). The similarity of all data shows that the microcalorimeters have membranes with very similar properties (composition, size, thickness).}
\label{figkappa}
\end{figure}

Different $\Delta T$ ranging from 2 to 5\% of $T_\circ$ are measured. $\kappa (T_\circ,\Delta T)$ is then fitted with a fourth order polynomial in $T_\circ$ and $\Delta T$ from the coefficients of which the thermal conductance $\kappa (T)$ is calculated. The typical error of the least square fit is $1-1.5\%$. For the sake of a comparison of the raw data, we present in Figure \ref{figkappa} the measured conductance $\kappa (T_\circ,\Delta T) \approx \kappa(T_\circ+\Delta T/2)$,  which is a good approximation for small $\Delta T$. 
The values of $\kappa (T_\circ,\Delta T)$ for the four microcalorimeters are scattered by less than 3 \%. This shows that basic parameters of the membrane such as the thickness, size and composition are very close in the four microcalorimeters. This is a very important conclusion that allows us to calculate the specific heat of the sample films from the difference $c_{sam}=c_{tot}-c_{add}$, where $c_{add}$ is the addenda heat capacity of the microcalorimeter with the conduction layer only.
 
\section{Results}

\begin{figure}
\centerline{\epsfbox{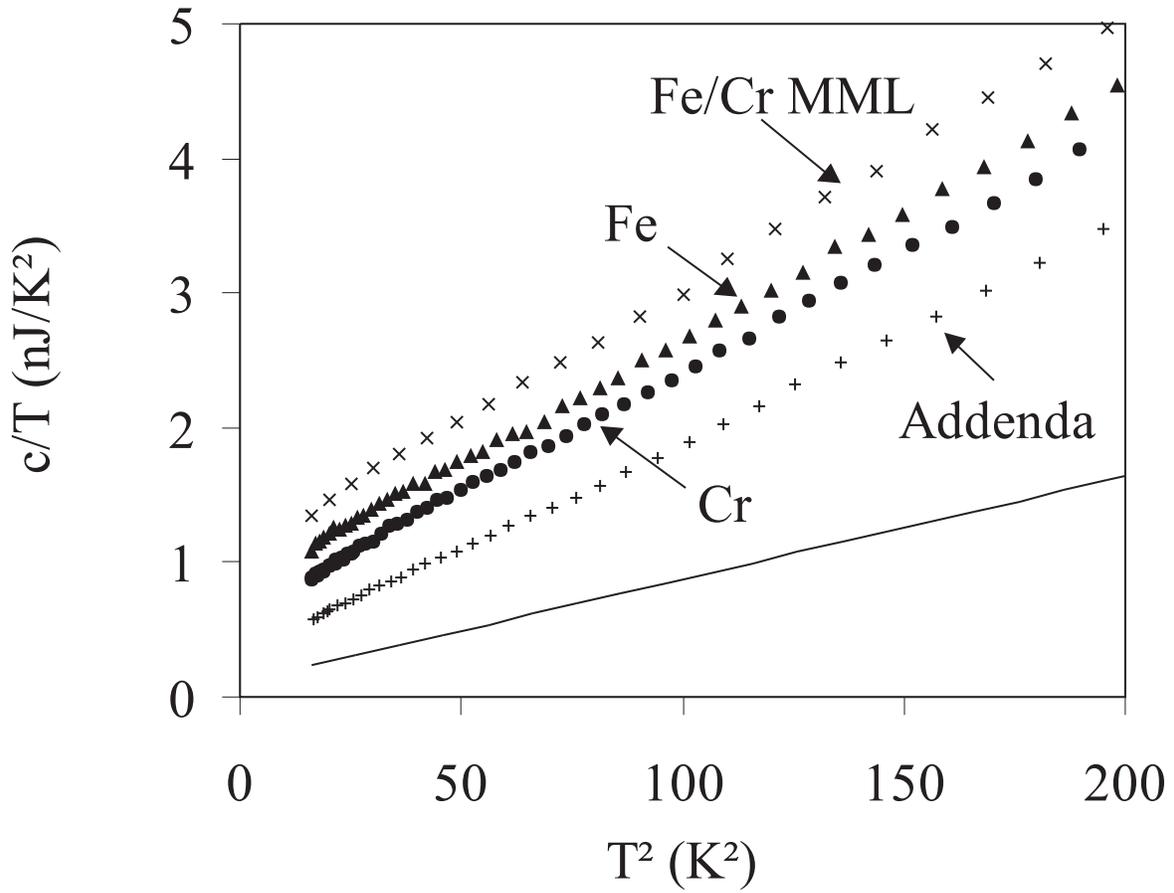}}
\caption{Total heat capacities of the addenda and Cr ($\bullet$), Fe ($\triangle$) and Fe/Cr samples ($\times$). Solid line: calculated contribution of the Cu conduction layer ($1810\;\AA$ thick) assuming bulk values $\gamma = 0.69 \; mJ/K^2mol$ and $\Theta_D=347\;K$.}
\label{figheatcapa}
\end{figure}

The total heat capacities $c_{tot}$ were calculated using the experimental values of $\tau_e$ and the fitted conductance $\kappa(T)$ and Equation \ref {one}. $c_{tot}$ was measured from $4\;K$ to $30\;K$ for the Fe and Cr samples and up to room temperature for the Fe/Cr MML and the addenda sample. $c_{tot}$ values are presented in Figure \ref {figheatcapa} in the temperature range $4\!<\!T\!<\!14\;K$ using the usual $c/T\;vs\;T^2$ representation. We report on the same plot an estimate of the contribution of the Cu layer calculated from the bulk coefficients (electronic coefficient $\gamma = 0.69 \; mJ/K^2mol$ and Debye temperature $\Theta_D=347\;K$ \cite{steward}). This contribution represents 40-50\% of the heat capacity of the addenda. The remaining part, which varies roughly as $\beta T^3$, is dominated by the $a\!-\!Si_{1-x}N_{x}$ membrane. Estimate of $\beta= 1.6\pm0.2\; \mu J/K^4g$ using the density $\rho_{SiN}=2.865\;g/cm^3$ is in good agreement with values measured by Zeller and Pohl for various glasses \cite{zeller}. Assuming a $Si_{0.5}N_{0.5}$ composition leads to $\Theta_D \approx 385\;K$. These values shouldn't be considered as the result of an actual fit of $c_{add}$. There is actually no way to separate the Cu and $Si_{0.5}N_{0.5}$ contributions using this set of data only.  The contributions of the Cr, Fe films and Fe/Cr multilayer to the total heat capacities are proportionally larger at low temperature (37\%, 48\% and 51\% respectively at $4\;K$), because the specific heat of the $a\!-\!Si_{1-x}N_x$ membrane decreases with $T^3$ in this temperature range. The total heat capacity is therefore dominated at the lowest temperatures by the linear terms of the specific heat of the metallic Cr, Fe and Fe/Cr films.

Since the thickness of the films and the molar densities of Fe and Cr are similar, important qualitative conclusions can already be drawn prior to any subtraction or normalization. The four curves plotted in Figure \ref{figheatcapa} appear as straight lines, which means that they can be described by the characteristic expression $c/T=a+bT^2$ observed in metals. No upturns are present down to $4\;K$. Furthermore, the curve of the MML lies above the curve of Fe and Cr. This indicates that the electronic term $a$ of the MML is not the simple mean of Fe and Cr weighted by their respective contribution to the mass of the MML as would be expected in a simple model. Finally, the slopes of the three upper curves  are similar, which implies lattice terms of similar magnitude in the three samples.

\begin{figure}
\centerline{\epsfbox{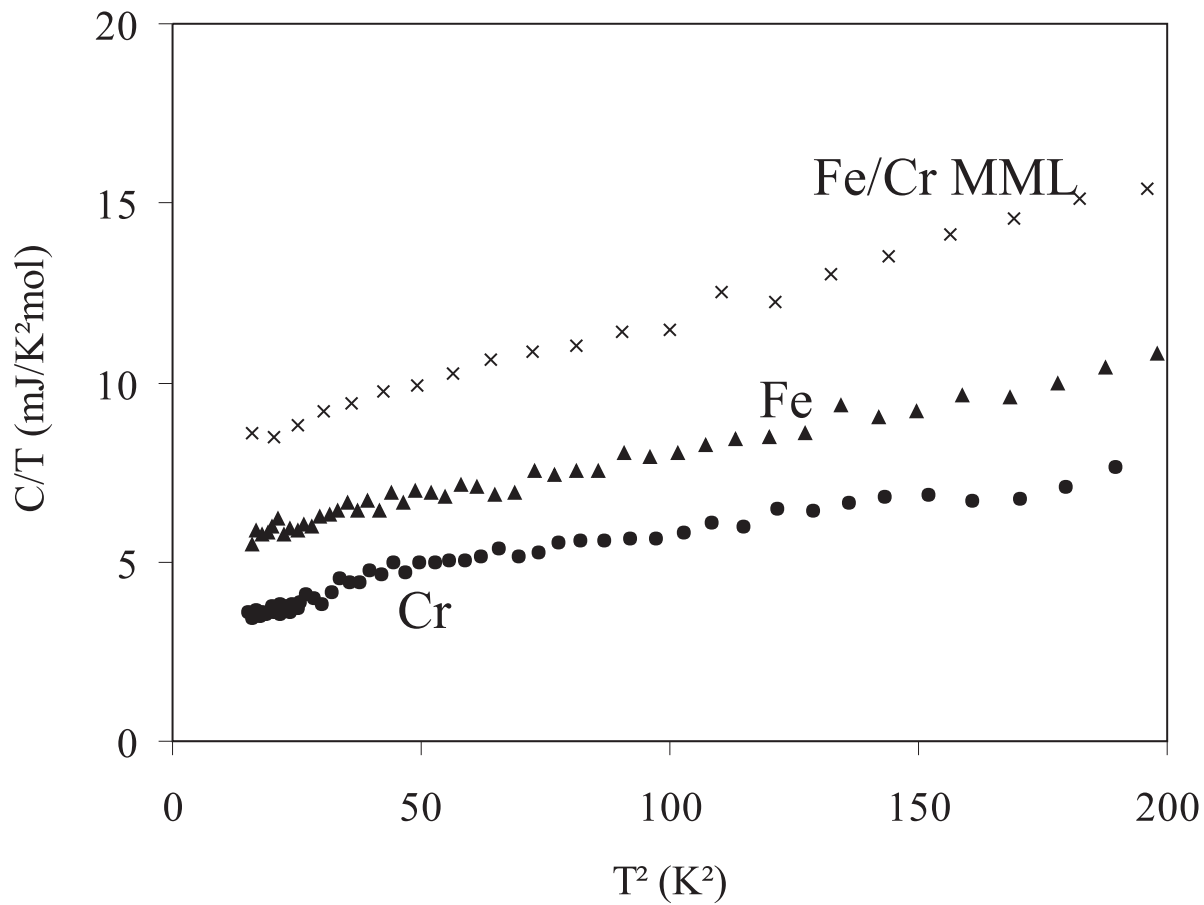}}
\caption{Specific heat of samples Cr, Fe and Fe/Cr samples (from bottom to top). The specific heat of the Fe/Cr MML has been normalized assuming a molar mass of Fe and Cr weighted by the ratio of Fe (66\%) and Cr (34\%) in the MML. As the molar masses of Fe and Cr differ by 2\% only, this normalisation does not depend critically on the ratio of Fe and Cr in the MML.}
\label{figsh}
\end{figure}

We now turn to a more quantitative analysis of the data. Figure \ref{figsh} shows the specific heat $C$ of the three samples in the temperature range $0<T<14\;K$. These three curves are the result of the normalized difference of the heat capacities of the sample microcalorimeters and the fitted heat capacity of the addenda. We assume that the heat capacity of the sample microcalorimeters $c_{tot}$ differ from $c_{add}$ only by the heat capacity of the Fe, Cr and Fe/Cr samples. This hypothesis is supported by the fact that all calorimeters were taken from the same wafer and had very similar values of $\kappa(T_\circ,\Delta T)$ (see Figure \ref{figkappa}). This is in agreement with the small variation of the thermal properties reported in the original paper on these microcalorimeters \cite{denlinger}. The Fe and Cr specific heat were normalized using the bulk values of the density ($7.87$ and $7.19\;g/cm^3$ respectively) and molar mass ($55.85$ and $52.00\;g/mol$ respectively). We used for the MML the sum of the total number of moles of Fe and Cr in the MML ($1.01\times10^{-7}$). Note that as the bulk Fe and Cr molar densities are the same within 2\% ($0.141$ and $0.138\; mol/cm^3$ respectively), this normalization does not depend critically on the respective values of the thickness of the layers in the MML. The error in the normalized specific heat $C$ has essentially two independent origins: first the experimental errors in $c_{tot}$, which are 2\% in total (see discussion above) and lead to a temperature dependent error in the difference (6\% at $4\;K$ and 12\% at $14\;K$ for the Fe/Cr MML) and second, the error introduced in the normalization by the uncertainty on the thickness and density of the films that is estimated conservatively to be 5\%.

We fitted the three curves with the usual expression of the specific heat of metals at low temperatures :

\begin{equation}
C/T=\gamma+\beta T^{2}
\label{three}
\end{equation}

where $\gamma$ is the electronic coefficient that is proportional to the effective mass of the electrons for a degenerate Fermi liquid. More precisely,

\begin{equation}
\gamma={k_B^{2}\pi^{2}\over 3}N(E_F)(1+\lambda)
\label{four}
\end{equation}

where $N(E_F)$ is the electronic DOS of the bare electrons and $(1+\lambda)$ is the mass enhancement factor caused, for example, by electron-phonon interactions. $\gamma$ is temperature independent for $k_BT<<E_F$ and $k_BT<<\hbar \omega_\circ$ ($E_F$ is the Fermi energy and  $\hbar \omega_\circ$ is the cut-off energy of the excitations responsible for the mass enhancement).
For a wide variety of compounds, the phonon contribution to the specific heat can be successfully described by a 3D linear Debye-like dispersion. At low temperature ($T<<\Theta_D$), this approximation leads to the second term in Equation \ref {three} with a coefficient $\beta$ that is:

\begin{equation}
\beta={12 \pi^{4} R \over 5 \Theta_D^{3}} \approx {1944 [J/Kmol] \over \Theta_D^{3}.}
\label{five}
\end{equation}

where $k_B\Theta_D$ is the cut-off energy of the phonons and $R$ is the molar gas constant.
$\gamma$ and $\Theta_D$ values obtained by fitting Equation \ref{three} to the $C$ data are summarized in Table \ref{table1} and compared with bulk values \cite{steward},\cite{landolt}. To avoid spurious values of $\Theta_D$ resulting from a fit made beyond the low temperature asymptote, we restrict the fit to the range $T\!<\!14\;K$, which is about 4\% of $\Theta_D$.

The presence of a magnetic term $\alpha T^{3/2}$ due to spin-wave excitations is expected in the specific heat of the Fe and Fe/Cr samples. Taking the bulk value $\alpha=0.028\;mJ/K^{5/2}mol$ measured in bulk Fe \cite{landolt}, results in a spin-wave contribution that is less than 1\% of the total specific heat in the range $4\!<\!T\!<\!14\;K$ and is therefore beyond our resolution. 

Spin-wave excitations in the Fe/Cr sample are more interesting because they are controlled by the coupling between the Fe layers. Theoretical calculation of the contribution of the spin-wave excitation to the specific heat of Co/Cu MML has been made by M\'elin and Fominaya \cite{melin}. This contribution is few percent of the total specific heat at low temperature and vanishes for a magnetic field larger than the saturation field. To investigate the presence of this contribution in our data, we measured the Fe/Cr MML in magnetic fields up to $8\;T$. The 0 of the microcalorimeters in magnetic fields is the subject of a separate paper \cite{zink}. The magnetic field {\bf B} was applied in the direction of the layers. We first checked that  $\kappa(T_\circ,\Delta T)$ was independent of the magnetic field. We then recorded the variation of $\tau(B,T_\circ)$ for fixed temperature $T_\circ=4, 5, 7.5$ and $10\;K$. No variation of $\tau(B,T_\circ)$ was observed within the 1\% resolution of this measurement. The absence of this term is explained by the presence of a gap in the spin wave excitations that is present in zero magnetic field already and that reduces the thermally accessible excitations. We note that high resolution electrical resistivity measurements on similar samples showed no indication of a spin-wave contribution in the temperature range $10\!<\!T\!<1\!50\;K$, which is in agreement with our result \cite{almeida}.

\begin{figure}
\centerline{\epsfbox{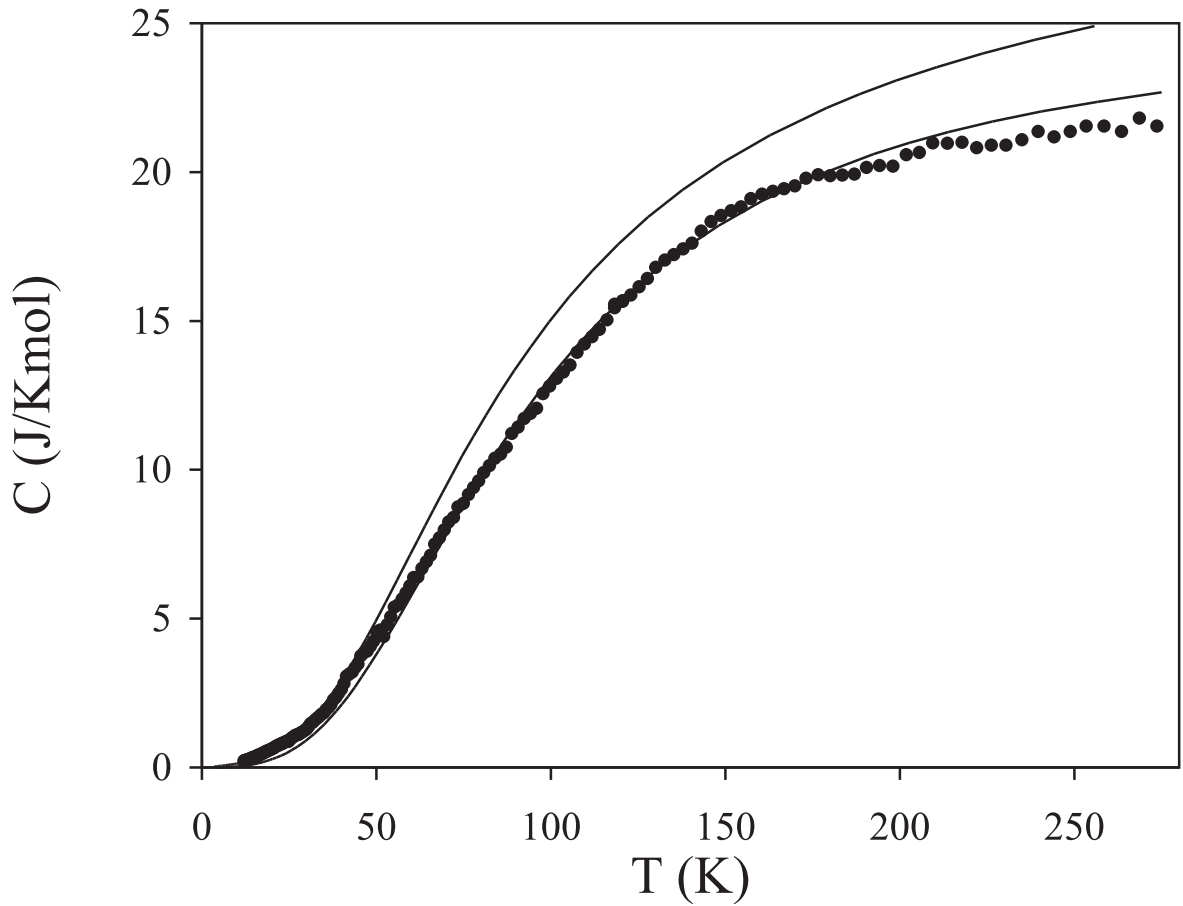}}
\caption{Specific heat of the Fe/Cr MML sample up to room temperature. Corrections due to thermal expansion are less than 1\% and are neglected (see text). Solid lines are calculated contribution from electrons and phonons in the Debye approximation. The upper curve is plotted using the low temperature parameters $\gamma = 8.7 \; mJ/K^2mol$ and $\Theta_D=358\;K$. The lower curve is a high temperature limit ($\gamma = 1 \; mJ/K^2mol$ and $\Theta_D=385\;K$). This plot shows that a temperature dependence of $\gamma$ and/or deviations from the Debye model are likely.}
\label{figshht}
\end{figure}

Finally, a measurement of the specific heat of the MML and the addenda was performed up to room temperature. The same procedure as that used for the low temperature was applied to extract the specific heat $C$ of the MML. Data are reported in Figure \ref{figshht}. To allow comparison with calculation made at constant volume($C_v$), $C$ measured at constant pressure ($C_p$) has to be corrected by a factor $(1+\beta^\circ \gamma^\circ T)^{-1}$ where $\gamma^\circ$ is the Gr\"uneisen coefficient and $\beta^\circ$ the volume thermal expansion coefficient (not to be confused with the coefficients $\gamma$ and $\beta$ used in Equation \ref{three}). To the best of our knowledge no such data are available for Fe/Cr MML. We therefore estimate the correction using the bulk Fe ($\gamma^\circ=1.67$, $\beta^\circ=35.4\times10^{-6}\;K^{-1}$) and Cr ($\gamma^\circ=1$, $\beta^\circ=14.7\times10^{-6}\;K^{-1}$) values  \cite {barron} and assuming isotropic 0, which gives a correction of about 1\% at $300\;K$. As this value is much less than our experimental errors, we neglect this correction in the next discussion.
We compare in Figure \ref{figshht} the specific heat of the MML to a simple model where we assume that $C$ is the sum of an electronic term with temperature independent $\gamma$ and a phonon term calculated from the Debye approximation, which gives:

\begin{equation}
C=\gamma T + 9Nk_B(T/\Theta_D)^3f_D(T/\Theta_D)
\label{six}
\end{equation}

where $f_D(T/\Theta_D)$ is the Debye function \cite{ashcroft}. No single set of parameters $\gamma$ and $\Theta_D$ could fit the entire range. A good fit is obtained up to $30\;K$ using Equation \ref{six} with the low temperature parameters $\gamma = 8.7\;mJ/K^2mol$ and $\Theta_D=358\;K$. Above this temperature, the data deviate significantly from the low temperature curve. In the range $70<T<200K$, a reasonable agreement is found using the parameters $\gamma < 1\;mJ/K^2mol$ and $\Theta_D=385\;K$. Note that the uncertainty on these latter parameters is much larger than at low temperature because of the correlation between $\gamma$ and $\Theta_D$ in Equation \ref{six} and the smaller contribution of the sample to the total heat capacity. 
However, it is clear that the deviation to the low temperature fit is outside the error bars. This deviation can be understood in  this simple model as the result of an increase of $\Theta_D$ and a decrease of $\gamma$. We mention that this conclusion would still hold in the case where the thermal expansion coefficient $\beta^\circ$ is very different from the bulk value $\beta^\circ_{bulk}$: On one hand, if $\beta^\circ<<\beta^\circ_{bulk}$ then the correction on $C_p$ becomes negligible and the same coefficients would be obtained. In the opposite limit $\beta^\circ>>\beta^\circ_{bulk}$, the correction would result in $C_v$ data that are smaller than those reported in Figure \ref{figshht}. As a consequence, $\Theta_D$ would be larger. Therefore, $\Theta_D=385\;K$ can be considered as a lower limit of the high temperature $\Theta_D$ in the approximation where the bulk thermal expansion is used. 

\section{Discussion}
We first comment on the low temperature $\Theta_D$ values. In all three samples a clear reduction of $\Theta_D$ with respect to bulk values indicates a softening of the low energy ($E<1\;meV$) phonon modes.
We note that $\Theta_D$ of the Fe and Cr films are similar ($410\;K$) in spite of different bulk values ($460$ and $610\;K$ \cite{steward}).  $\Theta_D$ of the Fe/Cr MML is further reduced with respect to these values ($356\;K$).

Very few experimental data are available on $\Theta_D$ values for thin films. Specific heat measurements of $1\;\mu m$ thick $a\!-\!Mo_{x}Ge_{1-x}$ films showed that for $x=0$ and $x=1$ $\Theta_D$ is close to the bulk values \cite{mael}. The same group reported $\Theta_D$ values for Nb/Zr ML \cite{broussard}. In the limit of very thick layers, they observe that $\Theta_D$ is smaller than the value expected in a non interacting model, which is explained by these authors by the presence of NbZr alloy at the interfaces.
Recent development of inelastic nuclear $\gamma$-ray scattering (INRS) has provided access to phonon DOS measurements on films as thin as $100\;\AA$.  Measurements on sputtered Fe films with different thickness \cite{rohlsberger} and nanocrystalline films \cite{fultz} showed that an enhancement of the low energy phonon DOS correlates with increasing disorder. This enhancement of the low energy DOS would result in a lower $\Theta_D$ as observed in our measurements.
The physical origin of this phenomena is not understood in detail. A phenomenological model introducing a finite lifetime of the phonons is used by these authors to fit the DOS with relative success. A similar correlation between disorder and decrease of $\Theta_D$ can be deduced from our data if one considers the textured nature of the films and the additional disorder introduced in the MML by the interdiffusion and roughness of the interfaces. We note that our result is consistent with the softening of the phonon modes measured in metallic superlattices by Schuller et al. \cite{schuller90}. It is important to add that the low energy part ($E<10\;meV$) of the DOS obtained by INRS follows roughly the square dependence in energy of the 3D Debye model for a surprisingly wide variety of samples. This dependence is in agreement with the $T^3$ term observed in our data.

We now turn to the electronic contributions $\gamma$. The value for Fe is reasonably close to the bulk value. A large enhancement of $\gamma_{Cr}$ is observed with respect to the bulk value of Cr ($1.4\;mJ/K^2mol$ \cite{muller}). Pure bulk Cr is 0 by an incommensurate spin density wave (ISDW) magnetic order, which suppresses $\gamma$ below the value for non-magnetic Cr ($3-3.5\;mJ/K^2mol$) that is calculated from the extrapolation of $\gamma$ for various Cr alloys (\cite{muheim} or see Fig. 37 of Ref. \cite{fawcett88}). This reduction of $\gamma$ in the ordered phase is explained by the gap opened on the Fermi surface by condensation of electron-hole pairs. As a consequence of this gap, the low temperature electronic entropy is reduced. The higher value of $\gamma$ for Cr observed in our data indicates that the bulk ISDW magnetism is not present in our Cr film. Bulk ISDW is known to be modified in epitaxial Cr films depending on the thickness and capping layer as seen in neutron scattering data by B\"odeker et al. \cite{bodeker99}. Combining this effect with the high sensitivity of the ISDW to disorder \cite{fawcett94} gives a plausible explanation for the increase of $\gamma_{Cr}$. Note that the same effect is not expected in Fe films of the same quality because the ferromagnetism of Fe is more robust due to the more localized nature of the electrons responsible for the magnetism in this case. We emphasize that the measurement of the electronic term $\gamma$ is a very direct probe of the magnetism of Cr since itinerant antiferromagnetism is related to the appearance of a gap on the Fermi surface. Moreover, the difference in $\gamma$ between the ordered and non-magnetic phase leads to a significant variation of the low temperature specific heat (more than 50\% at $4\;K$). Magnetic or transport measurements give less stringent criteria for the presence or absence of an ordered magnetic state in Cr: as the susceptibility of Cr is largely dominated by core electrons, the contribution of the itinerant electrons is only 3 percent of the total susceptibility (see Figure 35 of Reference \cite{fawcett88}). In transport measurements, the variation of the resistivity in the itinerant antiferromagnetism is due to the decrease of the electron spin scattering, which is an indirect effect.

The large value obtained for the Fe/Cr MML $\gamma_{ML}=8.7\;mJ/K^2mol$ is striking and more puzzling. Assuming that each layer in the MML can be described by the bulk values of Fe and magnetic Cr and assuming perfect interfaces leads to $\gamma_{ML}=3.6\;mJ/K^2mol$. Using the more realistic non-magnetic value for Cr \cite{fullerton96} increases $\gamma_{ML}$ only slightly to $4.3\;mJ/K^2mol$, half of the measured value. The absence of electronic band structure calculation complicates the interpretation of this result: purely electronic effects, effective mass enhancement, presence of other phases with high electronic $\gamma$ and other type of excitations are hard to separate. 

In the absence of excitations other than electronic, Equation \ref{four} shows that an increase of $\gamma$ can result from an increase of either $\lambda$ or $N(E_F)$. Note that these two quantities are not independent as $\lambda$ is proportional to $N(E_F)$. As a consequence, an increase in $N(E_F)$ results in an increase of $\lambda$ whereas the opposite is not true. Since no magnetic field dependence was observed, $\lambda$ is attributed to the electron-phonon interaction only. Softening of the phonons as observed in the reduced $\Theta_D$ of the Fe/Cr MML is known to increase $\lambda$. This is observed for example in the comparison of the superconducting properties of $Nb$ and $Nb_{75}Zr_{25}$ \cite{wolf}. In a simple Debye approximation that is valid only at low temperature ($q \to 0$), $\lambda \propto (1/\Theta_D^2)$ \cite{smith}. Nevertheless, this effect cannot account for the totality of the increase of $\gamma$: even a 40\% decrease of $\Theta_D$ - which would increase $\lambda$ by a factor two with respect to the bulk values of Fe and Cr i.e. 0.2 \cite{baker} - only increases $\gamma$ by 20\%. 

Another possible cause for the increase in $\gamma _ {ML}$ is the presence of an $Fe_{1-x}Cr_{x}$ alloy phase at the interfaces. Due to the very close atomic and electronic properties of Fe and Cr atoms and the continuous solubility of the alloy, interdiffusion at the interfaces of Fe/Cr MML is possible. Quantitative analysis of the interface roughness of Fe/Cr MML deposited under similar conditions has been made by Gomez et al. \cite{gomez} using energy-filtered transmission microscopy images. Using these data, we estimate the interdiffusion length as the standard deviation of the profile for the smallest window width, which is about $4.5\AA$. Values of $\gamma$ have been measured for bulk $Fe_{1-x}Cr_{x}$ alloy by Cheng et al. \cite{cheng}. Interestingly, the variation of $\gamma(x)$ for bulk samples is not monotonic and show a significant enhancement with respect to pure Fe and Cr values for $0.1\!<\!x\!<\!0.4$. This effect could not be reproduced in electronic band structure calculations by Kulikov and Demangeat \cite{kulikov} and Steward and Ruvalds \cite{steward} probably because many body effects at the origin of the SDW in Cr and the competing ferromagnetic order of the Fe spins are not taken into account in these calculations \cite{ruvalds01}. To estimate the effect of the presence of an Fe-Cr alloy on $\gamma _ {ML}$, we take an average value of $\gamma=7.2\; mJ/K^2mol$ calculated from the data of Cheng et al. distributed over $4.5\;\AA$ at the interfaces. For the rest of the MML, we use the bulk values of Fe and non-magnetic Cr. This simple model gives $\gamma=5.2\;mJ/K^2mol$. Combining this effect and the enhancement of $\lambda$ results in $\gamma \approx 6.2\;mJ/K^2mol$ still considerably below the measured $8.7\; mJ/K^2mol$. Note that the calculation of the contribution of the $Fe_{1-x}Cr_{x}$ alloy has been done using bulk values, which is questionable to describe few Angstr\"om  thick layers in proximity to a ferromagnetic Fe layer. In this context, local measurements of the electronic DOS in the perpendicular direction of the plane would be of great interest. 

We emphasize that an enhanced DOS at the interfaces of the MML whatever its origin, will play a role in the GMR: the combination of a large, partially polarized DOS $N(E_F)$ and scattering centers, both located at the interfaces, is one of the proposed mechanisms for GMR in MML. In calculations by Zahn et al. \cite{zahn} on CoCu MML, the enhancement of $N(E_F)$ at the interfaces is the consequence of the periodic potential of the layered structure that affects the electron states. This enhancement is essentially due to minority electrons of Co that are confined in the Co layer due to the mismatch of their band structure and that of Cu. These authors show that adding scattering centers at the interfaces causes the minority electrons to be more scattered than the majority electrons. As a consequence, the layer resolved GMR is  large at the interfaces. Considering that $Fe_{1-x}Cr_x$ alloy is polarized for $x>0.3$, is intrinsically disordered and has a large electronic DOS suggests that interdiffused interfaces play a similar role in Fe/Cr MML. This hypothesis is supported by the linear increase of the absolute GMR with the interface roughness observed recently by Santamaria et al. \cite{santamaria}.

\section{Conclusion}
We report in this paper specific heat measurements of two films of Fe and Cr and an Fe/Cr multilayer. A precise description of the method and techniques used to measure the specific heat of these $1000\;\AA$ thick films was given. In the temperature range $4<T<14\;K$, the specific heat of the three samples can be described with the usual expression expected for metals. A softening of the phonons is observed in all samples, which we suggest is a consequence of the disorder, in agreement with INRS data taken on similar samples. The electronic terms of the Fe and Cr are compatible with bulk values whereas the multilayer value is much larger than expected in any simple model. No dependence on magnetic field is seen (to better than 1\%) which rules out changes in electronic density of states as a unique source for GMR. A magnetic origin for the enhancement of $\gamma$ is excluded because of the absence of magnetic field dependence. Probable causes include an enhanced electron phonon interaction due to the softening of the phonons and the presence of an $Fe_xCr_{1-x}$ alloy at the interfaces with enhanced electronic density of states. We propose that the enhanced $\gamma$ of the multilayer plays a role in the magnetotransport, consistent with a model presented recently that emphasized the importance of interface states for GMR effects.

\acknowledgments

This work has been supported by the US Department of Energy. One of us (B.R.) thanks the Swiss National Fund for Scientific Research for financial support. We thank J. Ruvalds for enlightening comments and continuous interest in this research.

\break

\begin{table}
\caption{Result of the least square fit of Equation \ref{three} to the specific heat heat data of Figure \ref{figsh}. Bulk values are given in paranthesis.}
\begin{tabular}{llll}
  sample&$t\;[\AA]$&$\gamma \;[mJ/K^2mol]$&$\Theta_D\;[K]$\\
\tableline
 Cr											&  		1035		&$3.2\pm0.3$\ (1.4\tablenotemark[1], 3.5\tablenotemark[2]) & $415\pm13$ (610\tablenotemark[3])\\
 Fe											&  		1050		&$5.4\pm0.4$\ (4.95\tablenotemark[3]) 								&$415\pm13$ (460\tablenotemark[3])\\
 Fe/Cr MML								 & 		 1159		 &$8.7\pm0.7$\																								& $356\pm10$\\
\end{tabular}

\tablenotetext[1]{magnetic Cr, from Ref.\ \cite{fawcett88}.}
\tablenotetext[2]{non-magnetic Cr, from Ref.\ \cite{fawcett88}.}
\tablenotetext[3]{from Ref.\ \cite{landolt}.}
\label{table1}
\end{table}

\end{document}